\documentclass[11pt,a4paper]{article}
\usepackage{graphicx}
\usepackage{amsmath}
\usepackage{amssymb}
\usepackage{bm}
\usepackage{doi}
\usepackage[top=12mm,bottom=12mm,left=30mm,right=30mm,head=12mm,includeheadfoot]{geometry}
\numberwithin{equation}{section}
\usepackage{hyperref}

\usepackage{color}

\begin{document}

\begin{center}{\Large \textbf{
Massless Dirac fermions on a space-time lattice\smallskip\\
with a topologically protected Dirac cone}}\end{center}
\begin{center}
\textbf{A. Don\'{i}s Vela,$^{\mathbf{1}}$ M.J. Pacholski,$^{\mathbf{1}}$ G. Lemut,$^{\mathbf{1}}$ J. Tworzyd{\l}o,$^{\mathbf{2}}$\smallskip\\
 and C.W.J. Beenakker$^{\mathbf{1}}$}
\end{center}
\begin{center}
\textbf{1} Instituut-Lorentz, Universiteit Leiden, P.O. Box 9506,\\ 
2300 RA Leiden, The Netherlands\\
\textbf{2} Faculty of Physics, University of Warsaw, ul.\ Pasteura 5,\\
02--093 Warszawa, Poland
\end{center}
\begin{center}
(May 2022)
\end{center}
\section*{Abstract}
\textbf{
The symmetries that protect massless Dirac fermions from a gap opening may become ineffective if the Dirac equation is discretized in space and time, either because of scattering between multiple Dirac cones in the Brillouin zone (fermion doubling) or because of singularities at zone boundaries. Here we introduce an implementation of Dirac fermions on a space-time lattice that removes both obstructions. The quasi-energy band structure has a tangent dispersion with a single Dirac cone that cannot be gapped without breaking both time-reversal and chiral symmetries. We show that this topological protection is absent in the familiar single-cone discretization with a linear sawtooth dispersion, as a consequence of the fact that there the time-evolution operator is discontinuous at Brillouin zone boundaries. 
}

\section{Introduction}
\label{intro}

\subsection{Objective}

A three-dimensional (3D) topological insulator has gapless surface states with a conical dispersion \cite{Has10,Qi11}. This Dirac cone is protected by Kramers degeneracy, no perturbation that preserves time-reversal symmetry can gap it out --- provided that the top and bottom surfaces remain uncoupled, to prevent Dirac cones from annihilating pairwise \cite{Kan13}.

To study the dynamics of Dirac fermions on a computer, one needs to discretize the Dirac equation
\begin{equation}
i\hbar\left(\frac{\partial}{\partial t}+ v\,\bm{\sigma}\cdot\frac{\partial}{\partial \bm{r}}\right)\Psi(\bm{r},t)=V(\bm{r})\Psi(\bm{r},t)
\end{equation}
for the two-component spinor $\Psi(\bm{r},t)$ (with velocity $v$ and Pauli spin matrices $\sigma_\alpha$). The electrostatic potential $V$ preserves time-reversal symmetry, so one would expect the Dirac cone to remain gapless for any time-reversally invariant discretization scheme that avoids fermion doubling \cite{Nie81} (only zero-energy states at momentum $\bm{k}=0$).

The objective of our paper is, firstly, to demonstrate that this expectation is incorrect, it does not apply to the split-operator technique \cite{Bra99} for the discretization of the time-evolution operator, which is commonly used \cite{Kre04,Moc08,Fil12} because of its computational efficiency. Then, secondly, we will show how a ``drop-in'' modification of the algorithm can restore a gapless Dirac cone --- without reducing the computational efficiency (scaling as $N\ln N$ in the number of lattice sites).

\begin{figure}[tb]
\centerline{\includegraphics[width=0.5\linewidth]{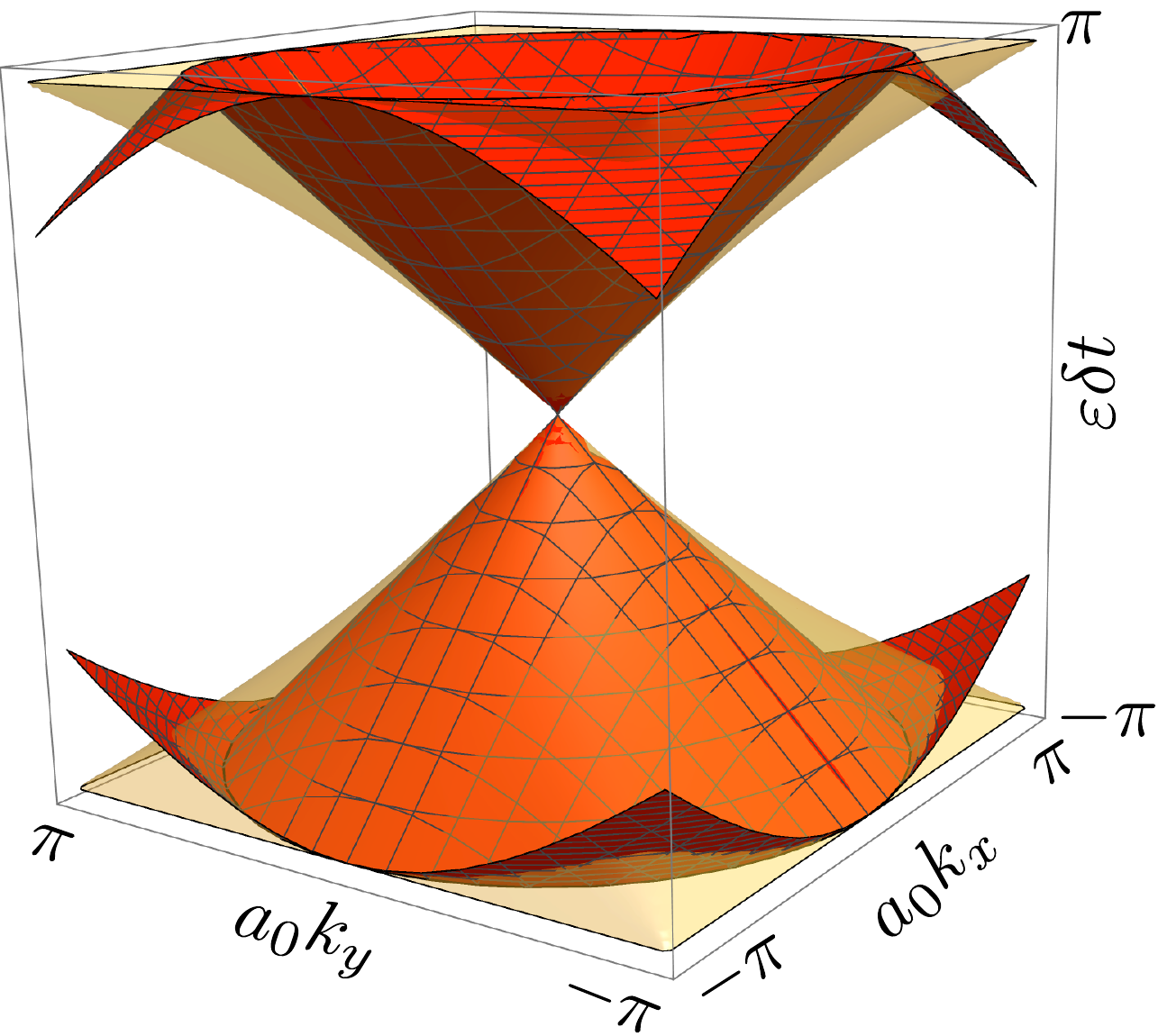}}
\caption{Quasi-energy bandstructure $\varepsilon(k_x,k_y)$ for the linear sawtooth dispersion (red) and for the tangent dispersion (yellow). The surfaces are computed, respectively, from the two equations $(\varepsilon\delta t+2\pi n)^2=(a_0k_x)^2+(a_0k_y)^2$, $n\in\mathbb{Z}$, and $\tan^2(\varepsilon\delta t/2)=\tan^2(a_0k_x/2)+\tan^2(a_0k_y/2)$. Only the first Brillouin zone is shown, the full bandstructure is periodic in momentum $k_\alpha$ with period $2\pi/a_0$ and periodic in quasi-energy $\varepsilon$ with period $2\pi/\delta t$. Near $\bm{k}=0$ both discretizations have the Dirac cone $\varepsilon^2=v^2(k_x^2+k_y^2)$ of the continuum limit, with velocity $v=a_0/\delta t$. A potential that varies rapidly on the scale of the lattice constant can gap out the Dirac cone for the linear sawtooth dispersion, but not for the tangent dispersion.
}
\label{fig_BZdispersion}
\end{figure}

We consider a 2+1-dimensional space-time lattice with lattice constants $a_0$ in space and $\delta t$ in time. In the split-operator technique the derivative operator $d/dx$ is evaluated in momentum representation as the linear function $k$ in the first Brillouin zone $|k|<\pi/a_0$ --- periodically repeated as a sawtooth for larger momenta. The drop-in modification that we propose is to replace $k$ by $(2/a_0)\tan (a_0k/2)$. The computational efficiency of the algorithm is not compromised, but the effect on the quasi-energy--momentum band structure is crucially important: While the linear sawtooth dispersion introduces discontinuous derivatives at Brillouin zone boundaries, the tangent dispersion produces a smooth band structure, see Fig.\ \ref{fig_BZdispersion}. As we will show, a potential that varies rapidly on the scale of $a_0$ is able to gap out the Dirac cone in the former case but not in the latter case.

By way of introduction, before we embark on the space-time discretization, we first discuss the simpler time-independent problem, when only space is discretized.  

\subsection{Time-independent problem}

\begin{figure}[tb]
\centerline{\includegraphics[width=0.5\linewidth]{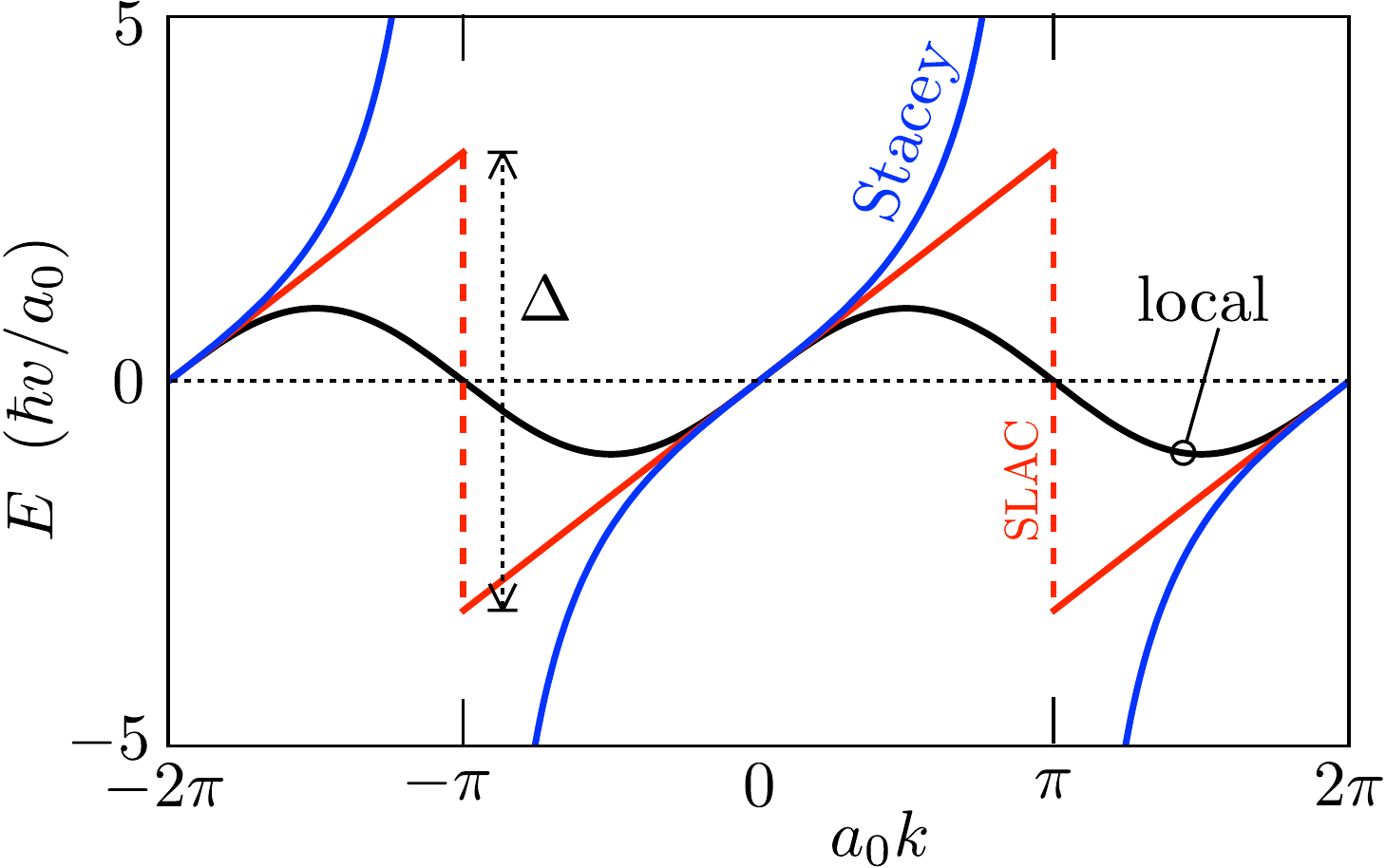}}
\caption{Three 1D dispersion relations, corresponding to a local discretization of the derivative operator $d/dx$ (black curve) and to two alternative nonlocal discretizations (red and blue curves).
}
\label{fig_dispersion}
\end{figure}

Consider a one-dimensional (1D) lattice along the $x$-axis, and first take $V\equiv 0$. Different ways to discretize the derivative $d/dx$ will produce different energy-momentum dispersion relations $\pm E(k)$. (The $\pm$ sign distinguishes the chirality of the massless Dirac fermions, left-movers versus right-movers.) What all dispersions have in common is that they are periodic with period $2\pi/a_0$ and vanish linearly at $k=0$. We compare three alternatives, see Fig.\ \ref{fig_dispersion}.

\begin{figure}[tb]
\centerline{\includegraphics[width=0.5\linewidth]{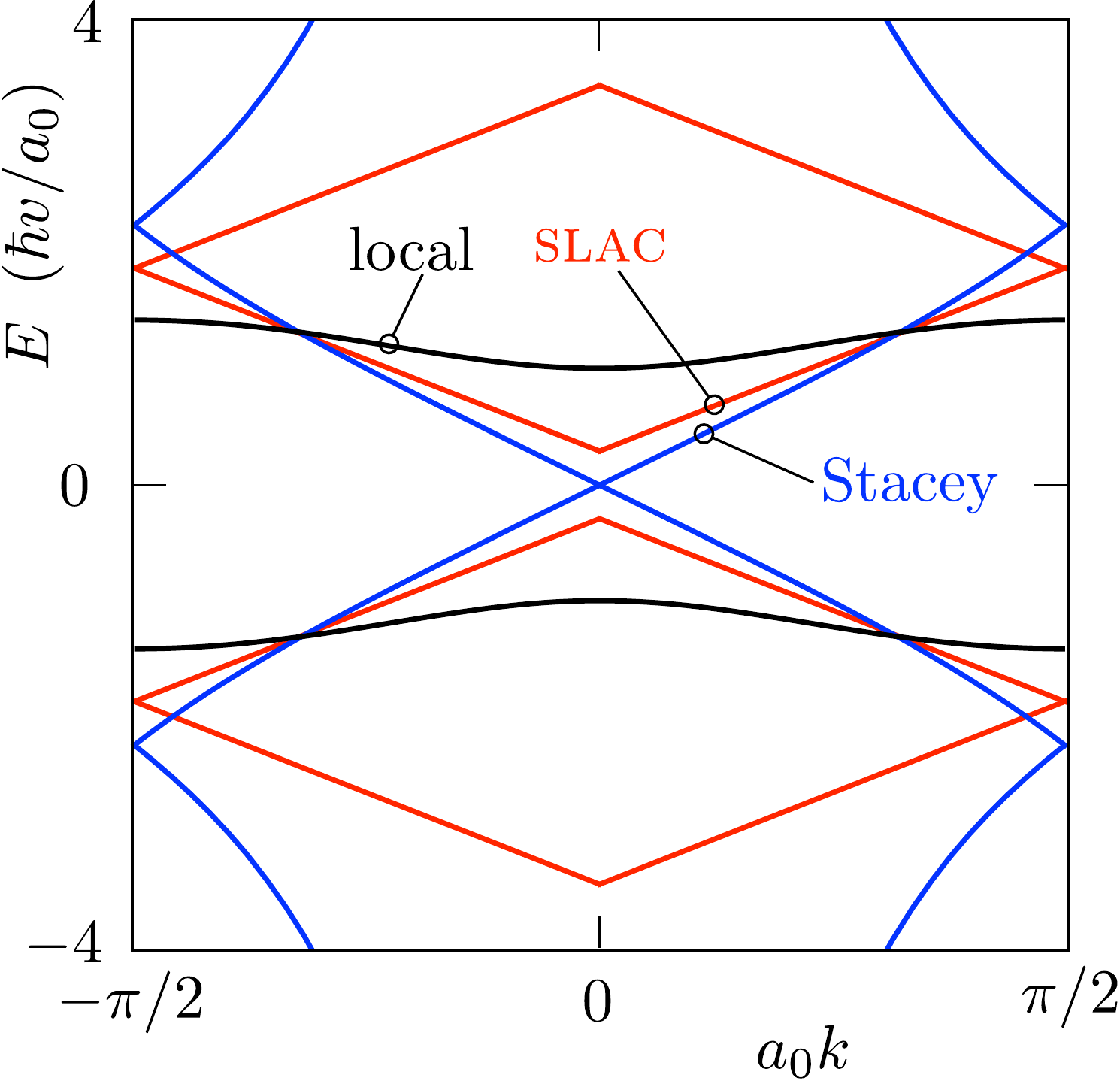}}
\caption{Band structure for three different spatial discretizations of the 1D Dirac Hamiltonian, with a staggered potential equal to $\pm 2\hbar v/a_0$ on even and odd-numbered lattice sites. The curves are computed from Eq.\ \eqref{EVresult}, with $E(k)$ given by Eqs.\ \eqref{Elocal}, \eqref{Ekslac}, and \eqref{Ekstacey} for the three discretizations. A gap opens at $k=0$ for the local discretization and for the {\sc slac} discretization, but not for the Stacey discretization. 
}
\label{fig_slacstaceylocal}
\end{figure}

The local discretization $df/dx\mapsto [f(x+a_0)-f(x-a_0)]/(2a_0)$ gives a sine dispersion 
\begin{equation}
E_{\rm local}(k)=\frac{\hbar v}{a_0}\sin(a_0 k),\label{Elocal}
\end{equation}
 which vanishes also at the boundary $|k|=\pi/a_0$ of the first Brillouin zone (fermion doubling). A nonlocal discretization, which couples $f(x)$ to distant lattice points, can remove the spurious Dirac cone at nonzero momentum. The socalled ``{\sc slac} discretization'' \cite{Dre76,Cos02} produces a dispersion relation that is strictly linear within the first Brillouin zone $|k|<\pi/a_0$. The dispersion has the $2\pi$-periodic sawtooth form\footnote{The function $\text{mod}\,(q,2\pi,-\pi)\equiv q -  2\pi\left\lfloor \frac{q+\pi}{2\pi}  \right\rfloor\in[-\pi,\pi)$ gives $q$ modulo $2\pi$ with an offset $-\pi$. (The floor function $\lfloor x\rfloor$ returns the greatest integer $\leq x$.) The mod function is discontinuous at $q=\pi$, jumping from $-\pi$ to $\pi$, we arbitrarily assign to $\text{mod}\,(\pi,2\pi,-\pi)$ the value of $-\pi$. The choice $\text{mod}\,(\pi,2\pi,-\pi)\equiv 0$ would produce in Fig.\ \ref{fig_slacstaceylocal} an isolated doubly degenerate state at $E=0=k$, disconnected from the {\sc slac} bands.}
\begin{equation}
E_{\rm SLAC}(k)=\frac{\hbar v}{a_0} \,\text{mod}\,(a_0 k,2\pi,-\pi).\label{Ekslac}
\end{equation}

Now apply the staggered potential $V(x)=V\cos(\pi x/a_0)$, switching from $+ V$ to $-V$ between even and odd-numbered lattice sites. This potential couples the states at $k$ and $k+\pi/a_0$, as described by the Hamiltonian
\begin{equation}
H_{V}(k)=\begin{pmatrix}
E(k)&V/2\\
V/2&E(k+\pi/a_0)
\end{pmatrix}.
\end{equation}
The Brillouin zone is halved to $|k|<\pi/2a_0$, with the band structure
\begin{equation}
 E_V(k)=\tfrac{1}{2}E(k)+\tfrac{1}{2}E(k+\pi/a_0)\pm\tfrac{1}{2}\sqrt{V^2+[E(k)-E(k+\pi/a_0)]^2}.\label{EVresult}
\end{equation} 
A gap opens in the Dirac cone for both the local and {\sc slac} discretizations, of size 
\begin{equation}
\delta E_{\rm local}=V,\;\;
\delta E_{\rm SLAC}=\frac{V^2a_0}{2\pi\hbar v}+{\cal O}(V^4).
\end{equation}

What we learn from this simple calculation is that removing the second cone at $|k|=\pi/a_0$ is not enough to protect the Dirac cone at $k=0$ from becoming gapped if the potential varies rapidly on the scale of the lattice constant. What happens is that the large gap $\Delta$ in the dispersion at $k=\pi/a_0$ is folded onto $k=0$ by the staggered potential, resulting in a minigap $\delta E=V^2/\Delta$ for $V\ll\Delta$. To avoid the gap opening we thus need a pole $\Delta\rightarrow\infty$ in the dispersion at the Brillouin zone boundary.

An alternative discretization due to Stacey \cite{Sta82} gives the dispersion
\begin{equation}
E(k)= (2\hbar v/a_0)\tan(a_0k/2),\label{Ekstacey}
\end{equation}
with a pole at $k=\pi/a_0$. And indeed, substitution of Eq.\ \eqref{Ekstacey} into Eq.\ \eqref{EVresult} shows that no gap opens at $k=0$ (see Fig.\ \ref{fig_slacstaceylocal}).

The merits of the Stacey discretization for the time-independent problem were studied in Refs.\ \cite{Two08} (at the level of the scattering matrix) and in Ref.\ \cite{Lem21} (at the level of the Hamiltonian). It was shown that the eigenvalue equation $H\Psi=E\Psi$ can be discretized into a \textit{generalized} eigenvalue problem ${\cal H}\Psi=E{\cal P}\Psi$ with \textit{local} Hermitian tight-binding operators on both sides of the equation.\footnote{The Stacey discretization is local in the sense that the operators ${\cal H}$ and ${\cal P}$ in the generalized eigenvalue problem ${\cal H}\Psi=E{\cal P}\Psi$ can be represented by \textit{sparse} Hermitian matrices. If we would write this as a strict (non-generalized) eigenvalue problem, ${\cal P}^{-1}{\cal H}\Psi=E\Psi$, we would find that the operator ${\cal P}^{-1}{\cal H}$ is nonlocal (it is not sparse). There is therefore no violation of the Nielsen-Ninomiya no-go theorem \cite{Nie81}, which only applies to strict eigenvalue problems.} Basically, a local formulation of the generalized eigenvalue problem is possible because tangent is the ratio of sine and cosine, which represent local tight-binding operators on a lattice. If all one would care about would be the presence of a pole in the dispersion at $k=\pi/a_0$, one could work with other functions than the tangent, but the tangent dispersion combines this property with the possibility of a local algorithm.

\subsection{Outline}

So much for the introduction to the time-independent discretization. In what follows we turn to the dynamical problem, by generalizing the approach of Refs.\ \cite{Sta82,Two08,Lem21} to the discretization of space and time. In the next section \ref{sec_stdiscrete} we show that the time discretization removes the pole in the tangent dispersion, which becomes a smooth function of momentum $\bm{k}$ and quasi-energy $\varepsilon$ (yellow bands in Fig.\ \ref{fig_BZdispersion}). In Sec.\ \ref{sec_stability} we then prove that the Dirac point remains gapless for any perturbation that preserves either time-reversal symmetry or chiral symmetry --- even if it varies rapidly on the scale of the lattice constant. 

In contrast, the quasi-energy bandstructure of the linear sawtooth dispersion has discontinuous derivatives at the Brillouin zone boundaries (red bands in Fig.\ \ref{fig_BZdispersion}). These spoil the protection of the Dirac cone, which is gapped by a staggered potential.

A key feature of the approach presented in Sec.\ \ref{sec_stdiscrete} is that it requires only a small modification of the usual split-operator technique, involving the replacement of the linear momentum operator appearing in the time-evolution operator by its tangent. Since this operator is evaluated in momentum representation, the replacement is immediate. It does not degrade the computational efficiency of the algorithm, which retains the favorable $N\ln N$ scaling in the number of lattice sites (limited only by the efficiency of the fast Fourier transform).

An alternative implementation which is fully in real space is possible, taking the form of an implicit finite-difference equation $A\Psi(t+\delta)=B\Psi(t)$ with sparse matrices $A$ and $B$. This formulation is a bit more cumbersome to explain, we present it an appendix.

\section{Space-time discretization without zone boundary discontinuities}
\label{sec_stdiscrete}

\subsection{Split-operator technique}
\label{sec_split}

The Dirac Hamiltonian
\begin{equation}
{\cal H}=v\bm{k}\cdot\bm{\sigma}+V(\bm{r})
\end{equation}
is the sum of a kinetic term that depends on momentum $\bm{k}$ and a potential term that depends on position $\bm{r}$. (We set $\hbar$ to unity.) The split-operator technique \cite{Bra99} separates these two terms in the time-evolution operator,
\begin{equation}
\begin{split}
&\Psi(t+\delta t)=e^{-i{\cal H}\delta t}\Psi(t),\;\;e^{-i{\cal H}\delta t}=U+{\cal O}(\delta t)^3,\\
&U=e^{-iV(\bm{r})\delta t/2}e^{-iv\delta t\,\bm{k}\cdot\bm{\sigma}}e^{-iV(\bm{r})\delta t/2},\label{Ukdef}
\end{split}
\end{equation}
with an error term that is of third order in the time slice $\delta t$ \cite{Suz90}. 

Space is discretized on a square or cubic lattice (lattice constant $a_0$ in each direction). The periodicity of the Brillouin zone is enforced by the substitution
\begin{equation}
\bm{k}\cdot\bm{\sigma}\mapsto a_0^{-1}\sum_\alpha \sigma_\alpha\,\text{mod}\, (a_0k_\alpha,2\pi,-\pi).\label{periodick}
\end{equation}
In 1D this is the linear sawtooth dispersion of Fig.\ \ref{fig_dispersion}, red curve. A discrete fast Fourier transform is inserted between the kinetic and potential terms, so that each is evaluated in the basis where the operators $\bm{k}$ and $\bm{r}$ are diagonal. The computational cost scales as $N\ln N$ for $N$ lattice sites.

\begin{figure}[tb]
\centerline{\includegraphics[width=0.5\linewidth]{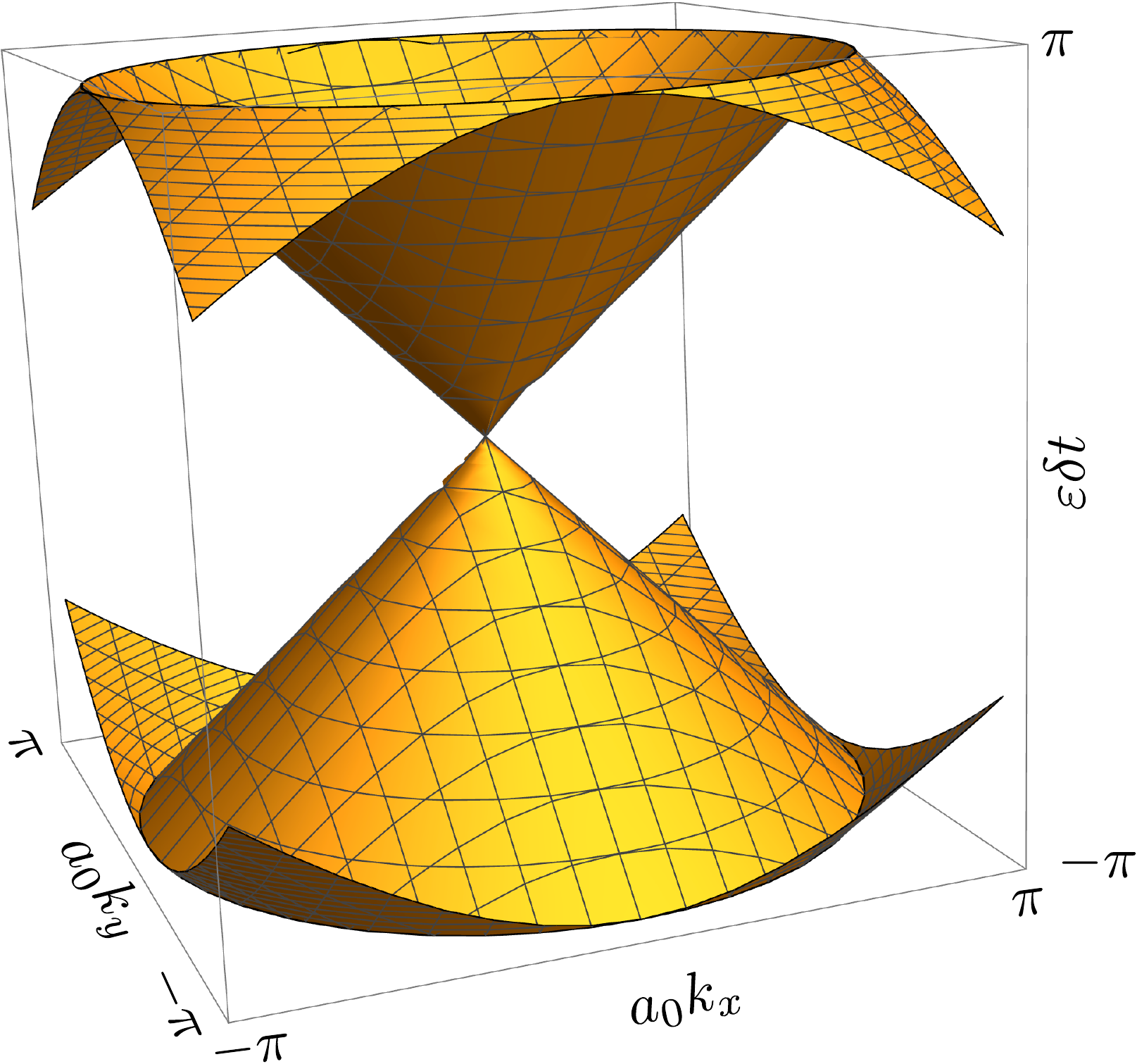}}
\caption{Momentum dependence of the quasi-energy for the free evolution operator $U$, given by Eq.\ \eqref{Ukdef} with $V=0$, computed from Eq.\ \eqref{dispersionk} in the 2+1 dimensional case. The space and time discretization units are related by $a_0=v\delta t$. Only the first Brillouin zone \eqref{calBdef} is shown.}
\label{fig_bandstructurek}
\end{figure}

\begin{figure}[tb]
\centerline{\includegraphics[width=0.5\linewidth]{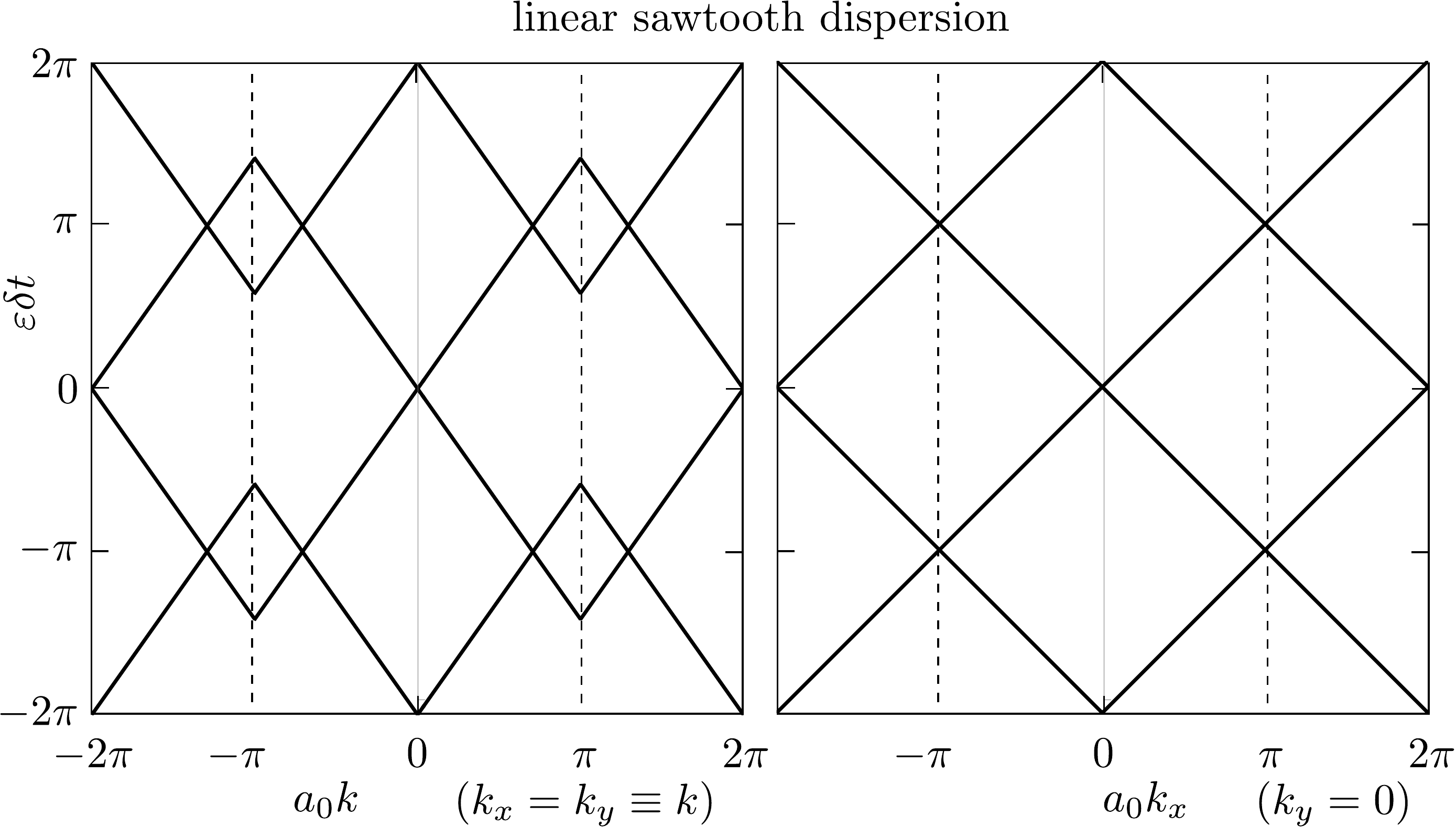}}
\caption{Cut through the bandstructure of Fig.\ \ref{fig_bandstructurek} along the line $k_x=k_y\equiv k$ (left panel) and along the $k_x$-axis (right panel). In the former direction the dispersion has a discontinuous slope at the Brillouin zone boundaries (dotted lines).}
\label{fig_dispersionk}
\end{figure}

The eigenvalues $e^{i\varepsilon\delta t}$ of the unitary operator $U$ define the quasi-energies $\varepsilon$ modulo $2\pi/\delta t$. For free motion, $V=0$, these are given by
\begin{equation}
(\varepsilon+2\pi n/\delta t)^2=v^2\sum_\alpha k_\alpha^2,\;\;n\in\mathbb{Z},\;\;|k_\alpha|<\pi/a_0.\label{dispersionk}
\end{equation}
The 2+1 dimensional band structure in the first Brillouin zone
\begin{equation}
{\cal B}=\{k_x,k_y,\varepsilon|-\pi<\varepsilon\delta t,k_x a_0,k_ya_0<\pi\}\label{calBdef}
\end{equation}
is plotted in Fig. \ref{fig_bandstructurek} for $v=a_0/\delta t$, when the dispersion is strictly linear along the $k_x$ and $k_y$-axes. (Alternatively, for $v=2^{-1/2}\, a_0/\delta t$ the dispersion is strictly linear along the diagonal lines $k_x=\pm k_y$, the corresponding plots are in App.\ \ref{app_sqrt2}.)

The band structure repeats periodically upon translation by $\pm 2\pi/a_0$ in the $k_x,k_y$ directions and by $\pm 2\pi/\delta t$ in the $\varepsilon$ direction. Upon crossing a zone boundary the dispersion has a discontinuous derivative, see Fig.\ \ref{fig_dispersionk}.

\subsection{Smooth zone boundary crossings}
\label{sec_removal}

To remove the discontinuity at the Brillouin zone boundary we modify the kinetic term in the evolution operator \eqref{Ukdef} in two ways: Firstly we approximate the exponent by a rational function (Cayley transform \cite{Wat00,Cha10}),
\begin{equation}
e^{-iv\delta t\,\bm{k}\cdot\bm{\sigma}}=\frac{1-\tfrac{1}{2}iv\delta t\,\bm{k}\cdot\bm{\sigma}}{1+\tfrac{1}{2}iv\delta t\,\bm{k}\cdot\bm{\sigma}}+{\cal O}(\delta t^3).
\end{equation}
The error of third order in the time slice is of the same order as the error in the operator splitting, Eq.\ \eqref{Ukdef}. 

Secondly we replace $k_\alpha$ by $(2/a_0)\tan (a_0k_\alpha/2)$, defining the modified evolution operator
\begin{subequations}
\label{Utildedef}
\begin{equation}
\tilde{U}=e^{-iV(\bm{r})\delta t/2}\frac{1-i(v\delta t/a_0)\,\sum_\alpha \sigma_\alpha\tan(a_0k_\alpha/2)}{1+i(v\delta t/a_0)\,\sum_\alpha \sigma_\alpha\tan(a_0k_\alpha/2)}e^{-iV(\bm{r})\delta t/2}.\label{Utildedefa}
\end{equation}
The inverse of the sum of Pauli matrices can be worked out, resulting in
\begin{equation}
\tilde{U}=e^{-iV(\bm{r})\delta t/2}\frac{[1-\sum_\alpha\chi^2(k_\alpha)]\sigma_0-2i\sum_\alpha\sigma_\alpha\chi(k_\alpha)}{1+\sum_\alpha\chi^2(k_\alpha)}
e^{-iV(\bm{r})\delta t/2}.\label{Utildedefb}
\end{equation}
\end{subequations}
We abbreviated $\chi(k)=(v\delta t/a_0)\tan(a_0k/2)$ and $\sigma_0$ is the $2\times 2$ unit matrix. This looks more complicated than Eq.\ \eqref{Ukdef}, but it can be computed equally efficiently since in both equations each operator is evaluated in the basis where it is diagonal.

The required periodicity when $k_\alpha\mapsto k_\alpha+2\pi/a_0$ is automatically ensured by the replacement of the linear momentum by the tangent, it does not need to be enforced by hand as in Eq.\ \eqref{periodick}. Although $\tan(a_0k_\alpha/2)$ has a pole when $k_\alpha=\pi/a_0$, this pole is removed in the evolution operator \eqref{Utildedef} ---  which has no singularity at the Brillouin zone boundaries.

\begin{figure}[tb]
\centerline{\includegraphics[width=0.5\linewidth]{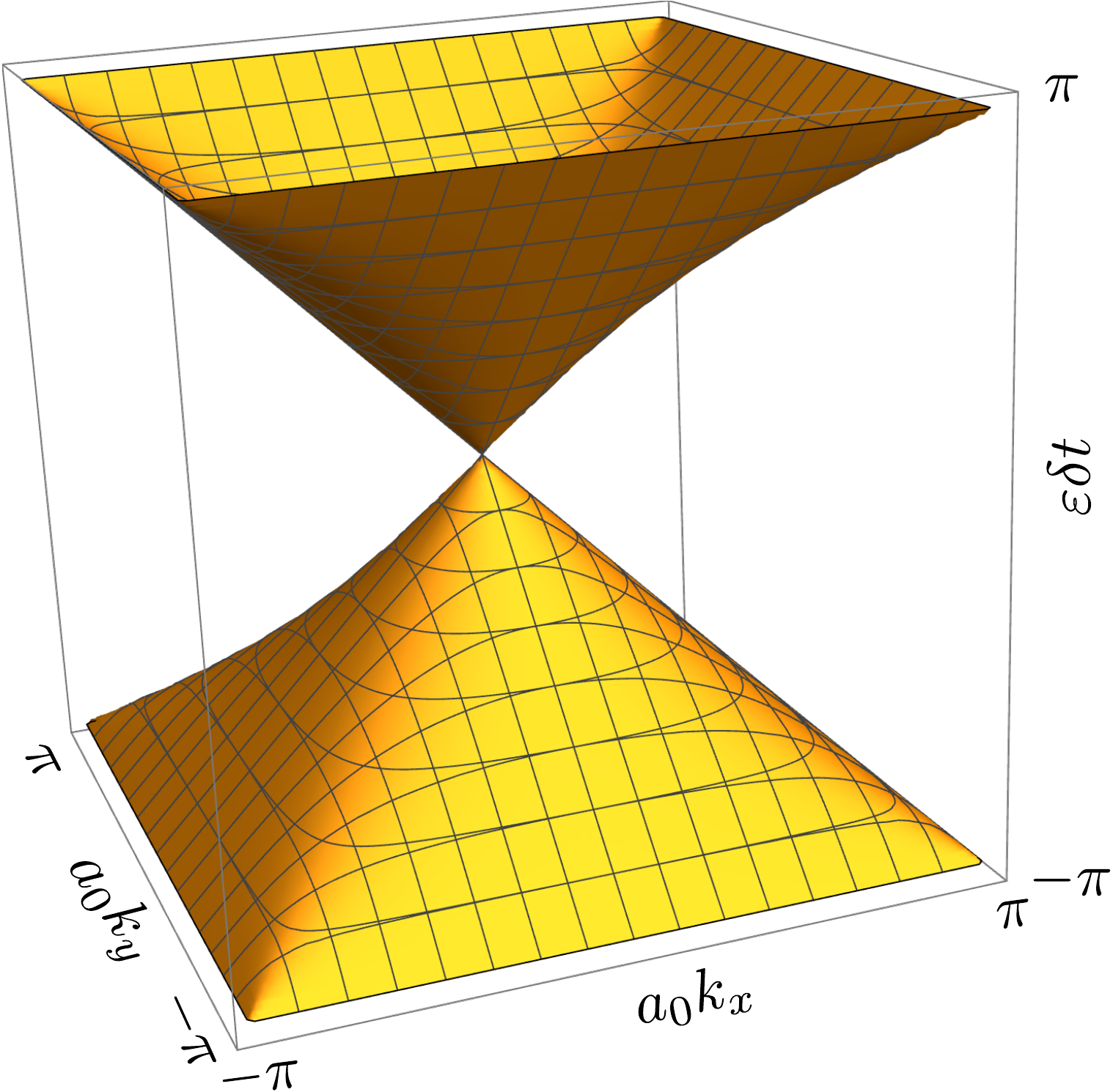}}
\caption{Same as Fig.\ \ref{fig_bandstructurek}, but now for the modified evolution operator \eqref{Utildedef} (with $v\delta t/a_0=1$).
}
\label{fig_bandstructuretan}
\end{figure}

\begin{figure}[tb]
\centerline{\includegraphics[width=0.5\linewidth]{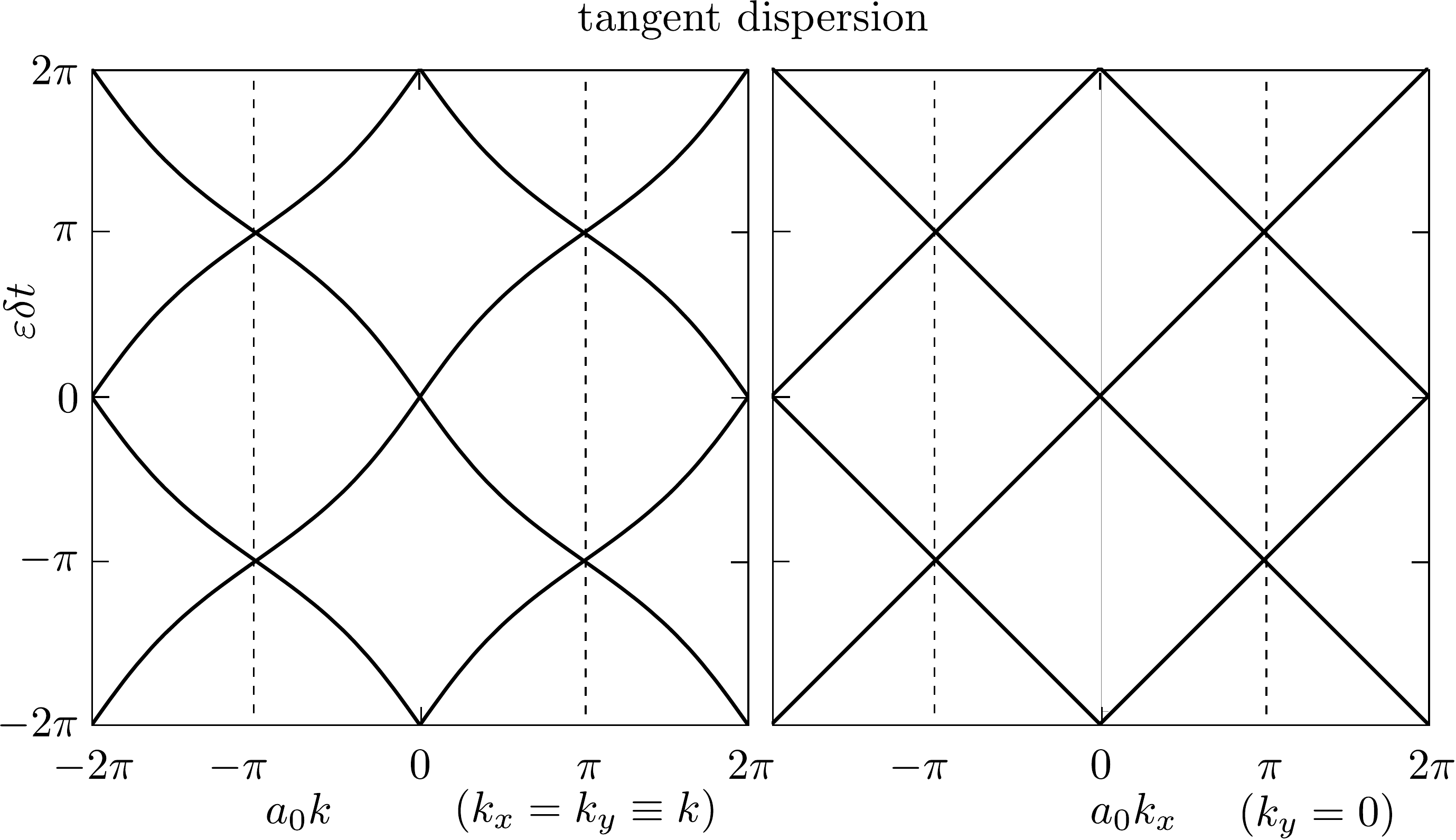}}
\caption{Cut through the bandstructure of Fig.\ \ref{fig_bandstructuretan} along the line $k_x=k_y\equiv k$ (left panel) and along the $k_x$-axis (right panel). In all directions the dispersion smoothly crosses the Brillouin zone boundaries (dotted lines).}
\label{fig_dispersiontan}
\end{figure}

The eigenvalues $e^{i\varepsilon\delta t}$ of $\tilde{U}$ for free motion, $V=0$, are given by
\begin{equation}
\tan^2(\varepsilon\delta t/2)=(v\delta t/a_0)^2\sum_\alpha \tan^2(a_0k_\alpha/2),\label{dispersiontan}
\end{equation}
plotted in Figs.\ \ref{fig_bandstructuretan} and \ref{fig_dispersiontan}. Comparison with Figs.\ \ref{fig_bandstructurek} and \ref{fig_dispersionk} shows that the zone boundaries are now joined smoothly. The dispersion is approximately linear near $\bm{k}=0$ and exactly linear along the lines $k_x=0$ and $k_y=0$ if we choose the discretization units such that $v=a_0/\delta t$. (See App.\ \ref{app_sqrt2} for the case $v=2^{-1/2}\,a_0/\delta t$, when the linear dispersion is along $k_x=\pm k_y$.)

\section{Stability of the Dirac point}
\label{sec_stability}

\subsection{Protection by time-reversal symmetry}
\label{TRS_protection}

The condition of time-reversal symmetry for the unitary evolution operator $U$ reads
\begin{equation}
\sigma_y U^\ast\sigma_y=U^{-1},\label{TRSeq}
\end{equation}
where the complex conjugation should be taken in the real space representation, when $\bm{k}=-i\nabla$ changes sign. The time-reversal operator, $\sigma_y\times\text{complex conjugation}$, squares to $-1$, so Kramers theorem applies: In the presence of a periodic potential $V$, when momentum $\bm{k}$ remains a good quantum number, the eigenvalues at $\bm{k}=0$ should be at least doubly degenerate.\footnote{Kramers theorem may be more familiar for a Hermitian operator, the proof for a unitary operator proceeds similarly: If $U\psi=e^{i\phi}\psi$ with $\phi\in\mathbb{R}$, and $\sigma_y U^\ast\sigma_y=U^{-1}$, then $U\sigma_y\psi^\ast=\sigma_y(\sigma_y U^\ast\sigma_y\psi)^\ast=\sigma_y(U^{-1}\psi)^\ast=e^{i\phi}\sigma_y\psi^\ast$, thus $\psi$ and $\sigma_y\psi^\ast$ are eigenstates of $U$ with the same eigenvalue. They cannot be linearly related, because if $\psi=\lambda\sigma_y\psi^\ast$ for some $\lambda\in\mathbb{C}$, then $\sigma_y\psi^\ast=-\lambda^\ast\psi=-|\lambda|^2\sigma_y\psi^\ast$, which is impossible for $\psi\neq 0$. Hence the eigenvalue $e^{i\phi}$ is at least doubly degenerate.}

\begin{figure}[tb]
\centerline{\includegraphics[width=0.4\linewidth]{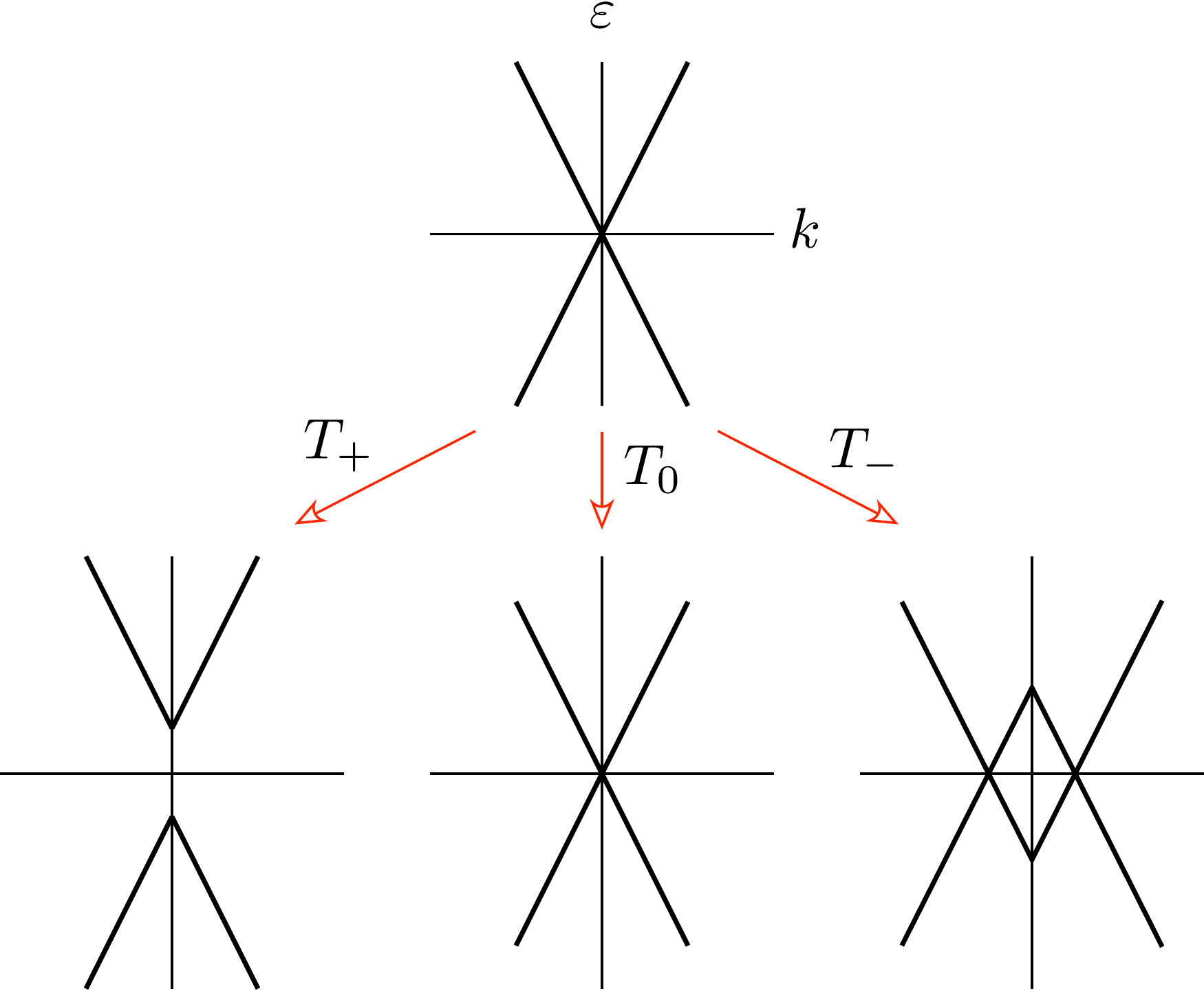}}
\caption{Top row: Dirac point in the quasi-energy dispersion $\varepsilon(k)$. Bottom row: Three topologically distinct modifications of the dispersion by the checkerboard potential. Only  the Dirac point preserving modification $T_0$ is allowed for an evolution operator that depends smoothly on momentum.
}
\label{fig_cutandsplit}
\end{figure}

Kramers degeneracy implies a band crossing at $\bm{k}=0$ --- provided that the bands depend smoothly on $\bm{k}$ --- hence this applies to the evolution operator $\tilde{U}$ for the tangent dispersion, but not to the operator ${U}$ for the linear sawtooth dispersion. We conclude that the Dirac point of $\tilde{U}$ is protected by time-reversal symmetry, while the Dirac point of $U$ is not.

\begin{figure}[tb]
\centerline{\includegraphics[width=0.8\linewidth]{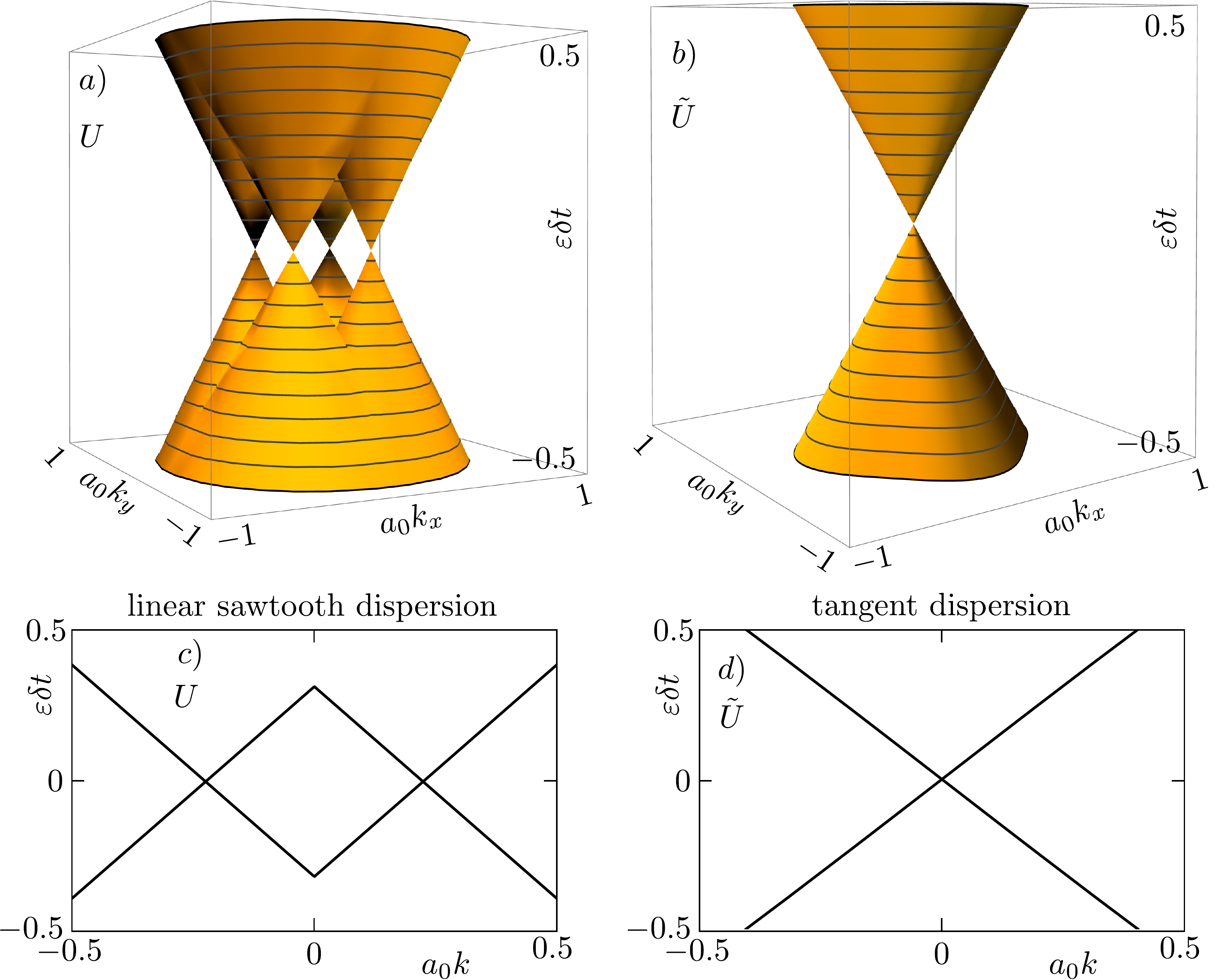}}
\caption{Quasi-energy bandstructure for the evolution operators $U$ (panels a,c) and $\tilde{U}$ (panels b,d), in the presence of the 2D checkerboard potential \eqref{checkerV} (for $V=2/\delta t=2\,v/a_0$). Panels c,d show a cut through the bandstructure for $k_x=k_y\equiv k$.}
\label{fig_checkerboard}
\end{figure}

We demonstrate this difference for the checkerboard potential
\begin{equation}
V(x,y)=V\cos[(\pi/a_0)(x+y)].\label{checkerV}
\end{equation}
(The calculation is described in App.\ \ref{app_checkerboard}.) In Fig.\ \ref{fig_cutandsplit} we show the three ways in which this potential can affect the Dirac point. The evolution operator $\tilde{U}$ shows the modification $T_0$, while $U$ shows $T_-$, see Fig.\ \ref{fig_checkerboard}. The other option $T_+$ appears in Fig.\ \ref{fig_slacstaceylocal} and in App.\ \ref{app_sqrt2}.

\subsection{Protection by chiral symmetry}
\label{chiral_protection}

Chiral symmetry of the evolution operator is expressed by
\begin{equation}
\sigma_z U\sigma_z=U^{-1}.\label{eq_chiral}
\end{equation}
Since $U^{-1}=U^\dagger$, this implies that $U$ can be decomposed in the block form
\begin{equation}
U=\begin{pmatrix}
A&B\\
-B^\dagger&C
\end{pmatrix},\;\;A=A^\dagger,\;\;C=C^\dagger.
\end{equation}

We consider a 2D periodic potential, so that momentum $\bm{k}=(k_x,k_y)$ is a good quantum number. The band structure has winding number \cite{Moc20}
\begin{equation}
W=\frac{1}{2\pi}\,\text{Im}\,\oint_\Gamma d\bm{k}\cdot\partial_{\bm{k}}\ln\det B(\bm{k})\in\mathbb{Z}
\end{equation}
along a contour $\Gamma$ in the Brillouin zone on which $\det B$ does not vanish.\footnote{One has $\det B\neq 0$ on $\Gamma$ if the quasi-energy $\varepsilon(\bm{k})$ does not cross 0 or $\pi$ on that contour \cite{Moc20}. This prevents us from extending the contour along the entire first Brillouin zone, when the winding number should vanish.}  This is a topological invariant, it cannot change in response to a continuous perturbation \cite{Ryu10}. A Dirac point within the contour is signaled by $W=\pm 1$. While pairs of Dirac points of opposite winding number can annihilate, a single Dirac point is protected by chiral symmetry --- provided that the evolution operator is continuous.

The 2D Dirac Hamiltonian has chiral symmetry when $V\equiv 0$. An in-plane magnetization
\begin{equation}
M(x,y)=\mu_x(x,y)\sigma_x+\mu_y(x,y)\sigma_y
\end{equation}
preserves the chiral symmetry. We are thus led to compare the two evolution operators
\begin{align}
&U=e^{-iM(x,y)\delta t/2}e^{-i(v\delta t/a_0)\,\sum_{\alpha=x,y}\sigma_\alpha\,\text{mod}\,(a_0k_\alpha,2\pi,-\pi)}e^{-iM(x,y)\delta t/2},\\
&\tilde{U}=e^{-iM(x,y)\delta t/2}\frac{1-i(v\delta t/a_0)\,\sum_{\alpha=x,y} \sigma_\alpha\tan(a_0k_\alpha/2)}{1+i(v\delta t/a_0)\,\sum_{\alpha=x,y} \sigma_\alpha\tan(a_0k_\alpha/2)}e^{-iM(x,y)\delta t/2}.
\end{align}
Both satisfy the chiral symmetry relation \eqref{eq_chiral}, $\tilde{U}$ is a continous function of $\bm{k}$ while $U$ is not.

\begin{figure}[tb]
\centerline{\includegraphics[width=0.8\linewidth]{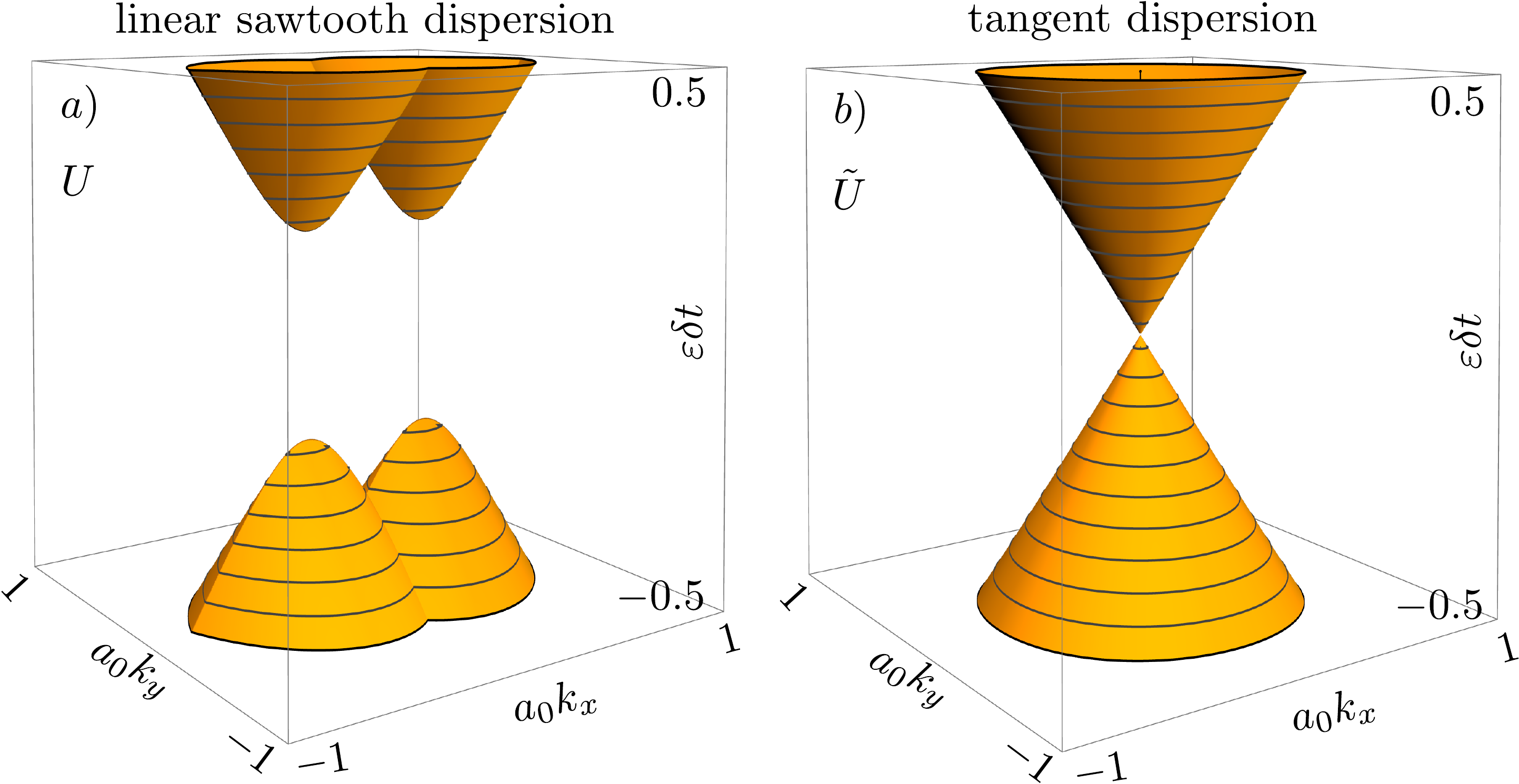}}
\caption{Quasi-energy bandstructure for the evolution operators $U$ (panel a) and $\tilde{U}$ (panel b), in the presence of the checkerboard magnetization \eqref{checkerM} (for $\mu=2/\delta t=2\,v/a_0$).}
\label{fig_chiral}
\end{figure}
 
The implication for the stability of the Dirac point is shown in Fig.\ \ref{fig_chiral}, where we compare the bandstructure in the presence of the checkerboard magnetization 
\begin{equation}
M(x,y)=\mu\sigma_x\cos[(\pi/a_0)(x+y)]\label{checkerM}
\end{equation}
(see App.\ \ref{app_checkerboard}). A gap opens for $U$ (linear sawtooth dispersion), while the Dirac point for $\tilde{U}$ (tangent dispersion) remains unaffected.
 
 \section{Conclusion}
 \label{conclude}
 
In conclusion, we have presented a method to cure a fundamental deficiency of the split-operator technique for the space-time discretization of the Dirac equation \cite{Bra99}. The linear sawtooth representation of the momentum operator preserves the time-reversal and chiral symmetries of the continuum limit, but it breaks the topological protection of the Dirac cone that these symmetries should provide. The deficiency originates from the discontinuity of the discretized time-evolution operator at the boundaries of the Brillouin zone. We have demonstrated the breakdown of the topological protection for a simple model: a periodic potential (or magnetization) on a 2D square lattice (lattice constant $a_0$) which couples the Dirac point at $k=0$ to the zone boundaries at $k=\pi/a_0$.

To restore the topological protection we modify the split-operator technique without compromising its computational efficiency, basically by replacing $a_0k$ in the evolution operator by $2\tan(a_0 k/2)$. Since the momentum operators are evaluated in the basis where they are diagonal, this is a ``drop-in'' replacement --- it does not degrade the $N\ln N$ efficiency of the split-operator algorithm.

One open problem of the split-operator technique that is not addressed by our modification is the difficulty to incorporate the vector potential in a gauge invariant way \cite{Roi13}. For that purpose it would be useful to formulate the split-operator technique fully in real space. This is done in Ref.\ \cite{Fil12} for the original approach with the linear sawtooth momentum operator. In App.\ \ref{app_realspace} we show that our tangent modification also allows for a real space formulation.

The availability of a single-cone discretization scheme which is efficient and which does not break the topological protection is a powerful tool for dynamical studies of massless Dirac fermions. One application to Klein tunneling has been published recently \cite{Don22}.

\section*{Acknowledgements}
C.B. received funding from the European Research Council (Advanced Grant 832256).\\
J.T. received funding from the National Science Centre, Poland, within the QuantERA II Programme that has received funding from the European Union's Horizon 2020 research and innovation programme under Grant Agreement Number 101017733, Project Registration Number 2021/03/Y/ST3/00191, acronym {\sc tobits}.

\appendix

\section{Bandstructures for $\mathbf{v=2^{-1/2}\,a_0/\delta t}$}
\label{app_sqrt2}

\begin{figure}[tb]
\centerline{\includegraphics[width=0.8\linewidth]{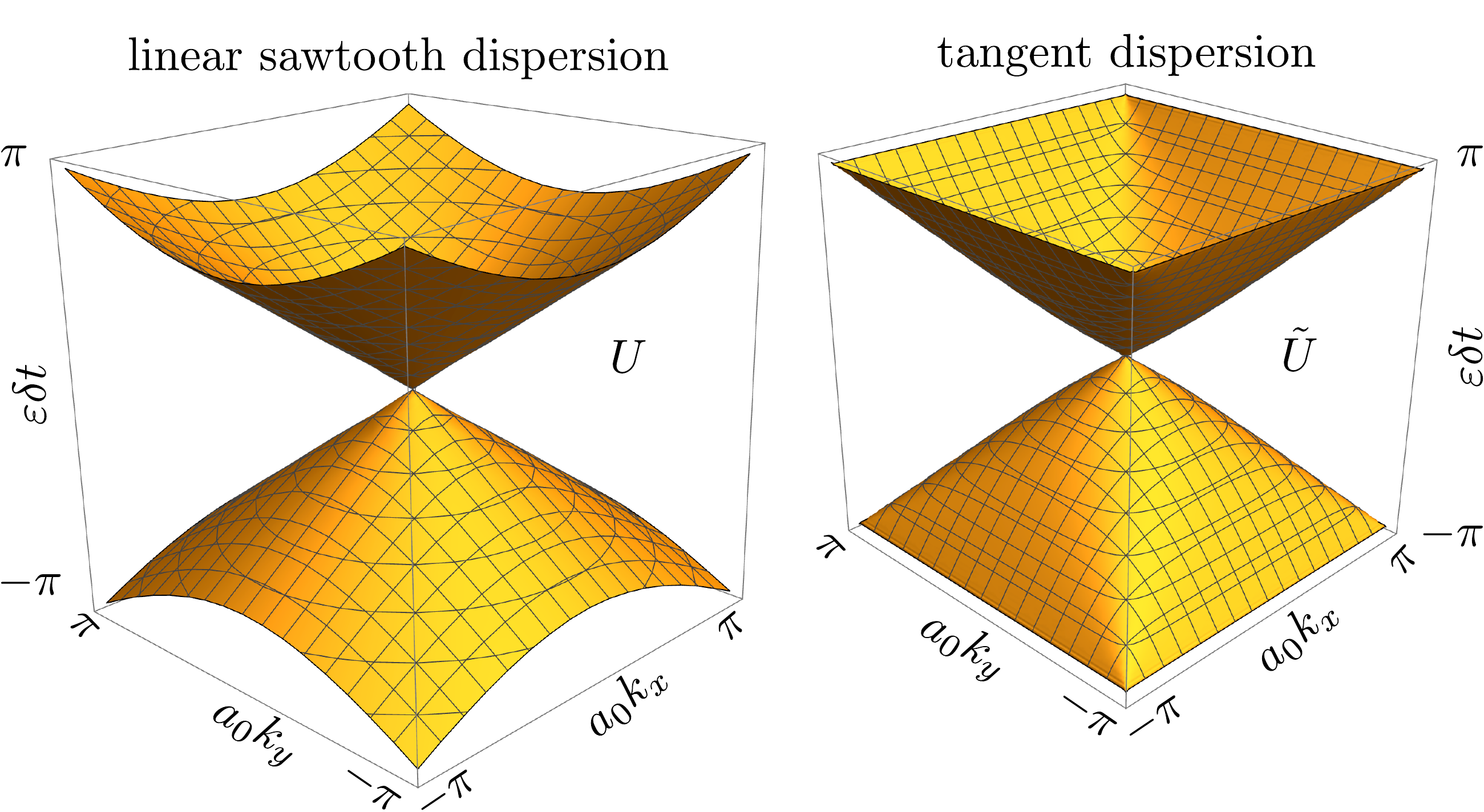}}
\caption{Free evolution ($V=0$) bandstructures of $U$ (left panel) and of $\tilde{U}$ (right panel), for $v=2^{-1/2}\,a_0/\delta t$.
}
\label{fig_multibandstructures}
\end{figure}

The bandstructures in the main text are for space-time discretization units such that $v=a_0/\delta t$, when the dispersion is strictly linear along the lines $k_x=0$ and $k_y=0$. Alternatively, one can have a strictly linear dispersion along the diagonals $k_x=\pm k_y$, by choosing $v=2^{-1/2}\,a_0/\delta t$. The bandstructures of $U$ and $\tilde{U}$ for free evolution are shown in Fig.\ \ref{fig_multibandstructures}.

\begin{figure}[tb]
\centerline{\includegraphics[width=0.8\linewidth]{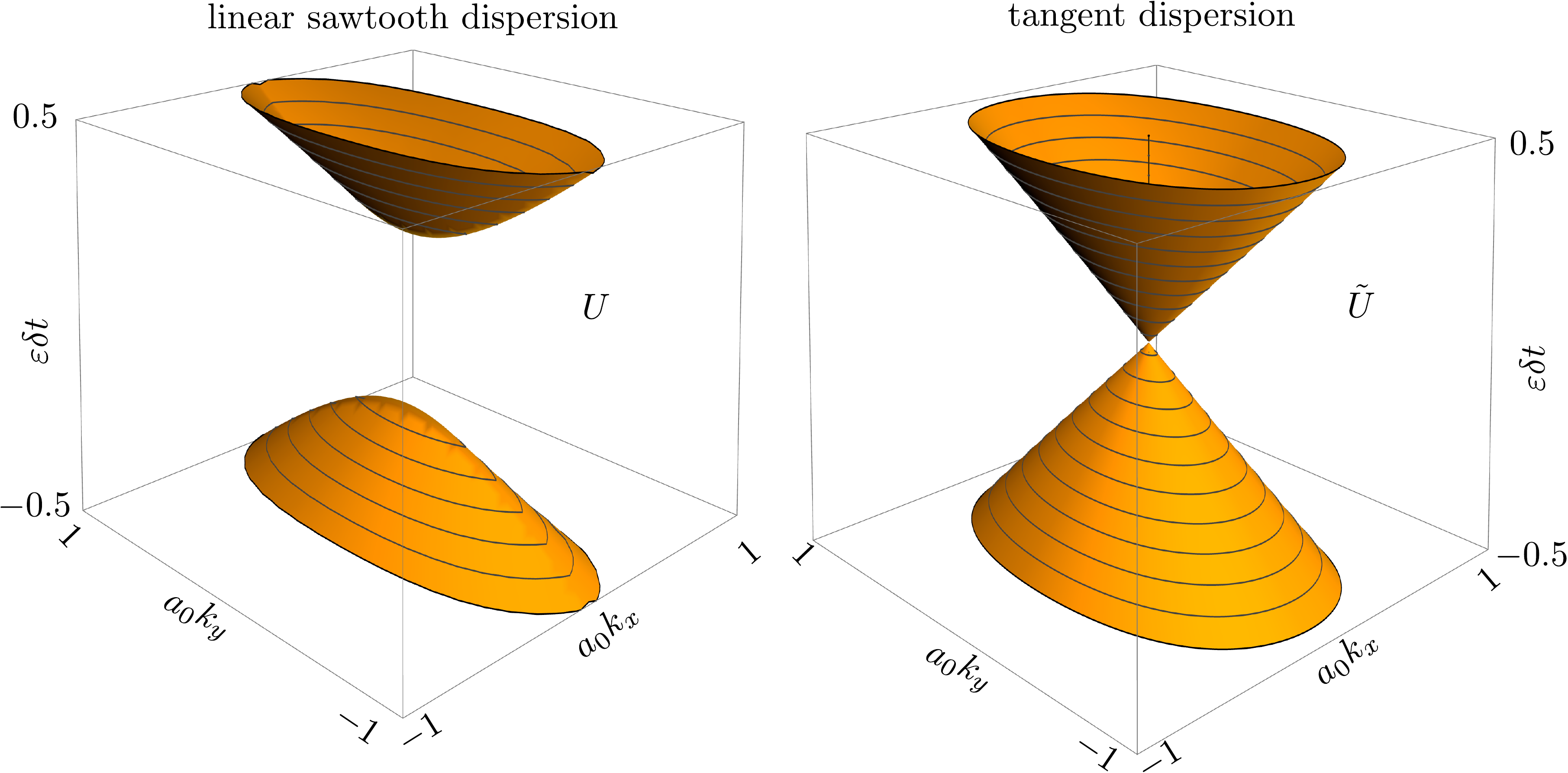}}
\caption{Same as Fig.\ \ref{fig_multibandstructures}, but now in the presence of the potential $V(x,y)=V\cos(\pi x/a_0)$ with $V=2\,\delta t$. The bandstructures for the staggered magnetization $M(x,y)=\mu\sigma_x\cos[(\pi/a_0)x]$ look very similar.}
\label{fig_potential}
\end{figure}

For $v=2^{-1/2}\,a_0/\delta t$ the checkerboard potential in the main text varies along the diagonals where $U$ is continuous, so it does not affect the Dirac point. Instead we choose here a staggered potential $V(x,y)=V\cos(\pi x/a_0)$ that varies along the $x$-axis. [In Eq. \eqref{Ucheckerboard} we thus replace $(k_x+\pi,k_y+\pi)$ by $(k_x+\pi,k_y)$.] The effect on $U$ is the $T_+$ gap-opening process of Fig.\ \ref{fig_cutandsplit}, while the Dirac point of $\tilde{U}$ is unaffected, see Fig.\ \ref{fig_potential}. We can also take the staggered magnetization $M(x,y)=\mu\sigma_x\cos[(\pi/a_0)x]$, with bandstructures very similar to those in Fig.\ \ref{fig_potential}.

\section{Bandstructure in the checkerboard potential}
\label{app_checkerboard}

In this appendix we choose $v=a_0/\delta t$ and set the discretization units $a_0,\delta t$ to unity. We compute the eigenvalues of the evolution operators $U$ and $\tilde{U}$ in the presence of the 2D checkerboard potential $V(x,y)=V\cos[\pi(x+y)]$. This potential couples states at $(k_x,k_y)$ and $(k_x+\pi,k_y+ \pi)$ with amplitude $V/2$.

We denote by $U_0(\bm{k})$ and $\tilde U_0(\bm{k})$ the free evolution operators, for $V=0$, given by
\begin{subequations}
\begin{align}
&U_0(\bm{k})=\exp\left(-i\textstyle{\sum_\alpha}\sigma_\alpha\,\text{mod}\,(k_\alpha,2\pi,-\pi)\right),\\
&\tilde{U}_0(\bm{k})=\frac{1-i\sum_\alpha\sigma_\alpha\tan(k_\alpha/2)}{1+i\sum_{\alpha}\sigma_\alpha\tan(k_\alpha/2)}.
\end{align}
\end{subequations}
The quasi-energies $e^{i\varepsilon}$ are the eigenvalues of the $4\times 4$ matrices
\begin{subequations}
\label{Ucheckerboard}
\begin{align}
&{\cal U}={\cal V}\begin{pmatrix}
U_0(k_x,k_y)&0\\
0&U_0(k_x+\pi,k_y+\pi)
\end{pmatrix}{\cal V},\\
&\tilde{\cal U}={\cal V}\begin{pmatrix}
\tilde{U}_0(k_x,k_y)&0\\
0&\tilde{U}_0(k_x+\pi,k_y+\pi)
\end{pmatrix}{\cal V}.
\end{align}
\end{subequations}
The $2\times 2$ blocks at $(k_x,k_y)$ and $(k_x+\pi,k_y+\pi)$ are coupled by the matrix
\begin{equation}
{\cal V}=\exp\left[-\frac{i}{2}\begin{pmatrix}
0&V/2\\
V/2&0
\end{pmatrix}\right]=\begin{pmatrix}
\cos(V/4)&-i\sin(V/4)\\
-i\sin(V/4)&\cos(V/4)
\end{pmatrix}.
\end{equation}
 Results for $V=2$ are plotted in Fig.\ \ref{fig_checkerboard}.
 
 For $\tilde{\cal U}$ the Dirac point at $\bm{k}=0$ is not affected by the checkerboard potential. In contrast, for ${\cal U}$ the $T_-$ modification of Fig.\ \ref{fig_cutandsplit} replaces the band crossing at $\bm{k}=0$ by four band crossings at $\pm(q,q)$ and $\pm(q,-q)$, with 
\begin{equation}
\cos \left(\frac{\pi -2 q}{\sqrt{2}}\right)=\cos \left(\frac{\pi }{\sqrt{2}}\right) \cos (V/2)\Rightarrow
q=0.067\,V^2+{\cal O}(V^4).
\end{equation}

The calculation for a checkerboard magnetization $M(x,y)=(\mu_x\sigma_x+\mu_y\sigma_y)\cos[\pi(x+y)]$ proceeds entirely similar, upon replacement of ${\cal V}$ by
\begin{equation}
{\cal M}=\exp\left[-\frac{i}{4}\begin{pmatrix}
0&\mu_x-i\mu_y\\
\mu_x+i\mu_y&0
\end{pmatrix}\right].
\end{equation}
The bandstructure for $\mu_x=2$, $\mu_y=0$ is shown in Fig.\ \ref{fig_chiral}. For evolution operator ${\cal U}$ the spectrum acquires a gap $\Delta\epsilon=0.095\,\mu_x^2+{\cal O}(\mu_x^4)$. For $\tilde{U}$ the Dirac cone remains gapless.

\section{Real-space formulation of the split-operator discretized evolution operator}
\label{app_realspace}

\subsection{Implicit finite-difference equation}
\label{implicitFDE}

The discretized Dirac equation for the tangent dispersion, $\Psi(t+\delta t)=\tilde{U}\Psi(t)$ with $\tilde{U}$ given by Eq.\ \eqref{Utildedef}, can be rewritten as a \textit{local} implicit finite-difference equation in real space --- without requiring a Fourier transform to momentum space.

We introduce the translation operator $r_\alpha\mapsto r_\alpha+a_0$ on a square or cubic lattice, given by $T_\alpha=e^{a_0\partial_\alpha}$, with $\partial_\alpha=\partial/\partial r_\alpha=ik_\alpha$. We note the identity
\begin{equation}
i\tan(a_0 k_\alpha/2)=\frac{T_\alpha-1}{T_\alpha+1}.
\end{equation}
The product operators
\begin{equation}
D_0=\tfrac{1}{4}\prod_\alpha(T_\alpha+1),\;\;D_\alpha=\tfrac{1}{2}(T_\alpha-1)\prod_{\alpha'\neq\alpha}(T_{\alpha'}+1)
\end{equation}
couple nearby sites on the lattice. 

The split-operator evolution equation
\begin{equation}
\Psi(t+\delta t)=e^{-iV(\bm{r})\delta t/2}\frac{1-i(v\delta t/a_0)\,\sum_\alpha \sigma_\alpha\tan(a_0k_\alpha/2)}{1+i(v\delta t/a_0)\,\sum_\alpha \sigma_\alpha\tan(a_0k_\alpha/2)}e^{-iV(\bm{r})\delta t/2}\Psi(t)
\end{equation}
can be rewritten identically in terms of these local operators,
\begin{equation}
\left(D_0+\frac{v\delta t}{2a_0}\sum_\alpha\sigma_\alpha D_\alpha\right)e^{iV(\bm{r})\delta t/2}\Psi(t+\delta t)=\left(D_0-\frac{v\delta t}{2a_0}\sum_\alpha\sigma_\alpha D_\alpha\right)e^{-iV(\bm{r})\delta t/2}\Psi(t).\label{secondevolution}
\end{equation}

The finite-difference equation \eqref{secondevolution} of the form $A\Psi(t+\delta t)=B\Psi(t)$ is called ``implicit'', because one needs to solve for the unknown $\Psi(t+\delta t)$ given the known $\Psi(t)$. The matrices $A$ and $B$ are both sparse, each of the $N$ sites on the 2D square lattice is only coupled to its four nearest neighbors. The method of nested dissection then allows for an efficient solution of the finite difference equation \cite{Geo83,Geo88,SuperLU}: There is an initial $N^{3/2}$ overhead from the LU decomposition of the matrix $A$, but subsequently the computational cost per time step scales as $N\ln N$ with the number of lattice sites, which is the same scaling as the split-operator algorithm. 

\subsection{Computational efficiency}
\label{sec_efficiency}

To check the efficiency of the discretization schemes we have calculated \cite{zenodo} the spreading of a wave packet in a 2D disordered lattice (of $M\times M$ sites, with periodic boundary conditions in $x$- and $y$-directions). We take a random potential $V(x,y)$ which varies independently on each of the $N=M^2$ sites, uniformly in the interval $(-0.5,0.5)\times\hbar v/a_0$. The initial state is
\begin{equation}
\Psi(x,y,0)=(4\pi w^2)^{-1/2}e^{ik_0x}e^{-(x^2+y^2)/2w^2}{1\choose 1},\label{wavepacket}
\end{equation}
with parameters $k_0=0.5/a_0$, $w=30\,a_0$. We follow the time evolution for $T=10^3$ time steps $\delta t=2^{-1/2}a_0/v$.

We compare the run time of the finite-difference code for a range of values of $N$, distinguishing the time $t_{\rm initial}$ spent on the initial LU decomposition from the run time $t_{\rm evolution}$ per time step needed for the subsequent evolution of the wave packet. (The full run time of the code is $t_{\rm initial} + Tt_{\rm evolution}$.)

\begin{figure}[tb]
\centerline{\includegraphics[width=0.9\linewidth]{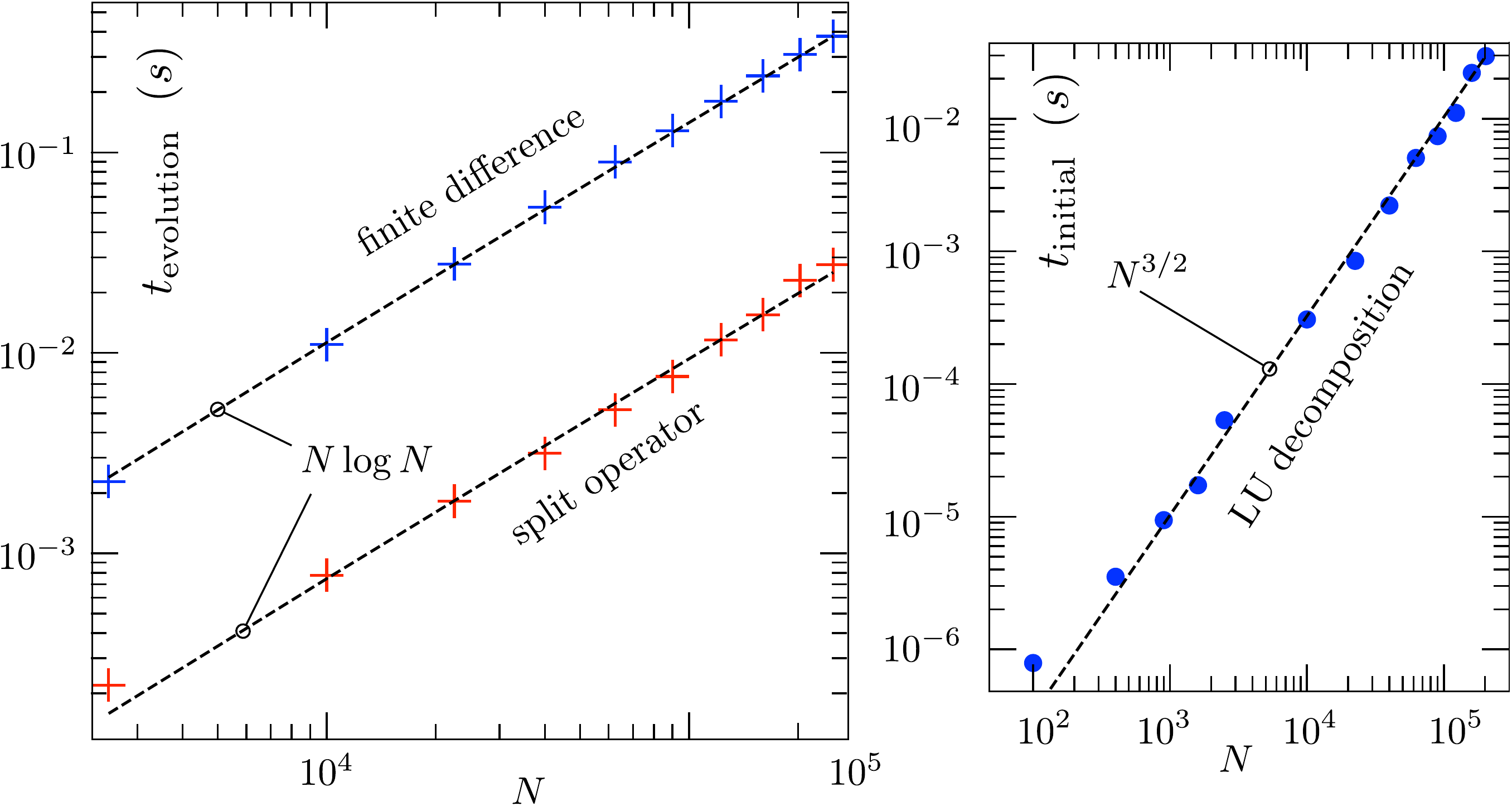}}
\caption{Demonstration of the favorable $N\ln N$ scaling with the number $N$ of lattice points of the single-cone discretization scheme with the tangent dispersion. The plot at the left shows the run time $t_{\rm evolution}$ per time step for the evolution of the wave packet \eqref{wavepacket} through a disordered 2D system: red symbols for the split-operator approach, blue symbols for the implicit finite-difference approach. The latter approach has an initial overhead $t_{\rm initial}\propto N^{3/2}$ from the LU decomposition, shown in the right plot.
}
\label{fig_efficiency}
\end{figure}

The data shown in Fig.\ \ref{fig_efficiency} is consistent with the expected scaling $t_{\rm initial}\propto N^{3/2}$ and $t_{\rm evolution}\propto N\ln N$. The storage requirements also scale as $N\ln N$, governed by the number of nonzero matrix elements in the LU decomposition.

We also show in the same plot the run time per time step for the split-operator algorithm. There is no initialization overhead in that case, the full run time is set by the $N\ln N$ cost of the fast Fourier transform.

\end{document}